\documentclass[12pt]{article}
\usepackage{graphicx}
\def\beq{\begin{equation}}
\def\eeq{\end{equation}}
\def\bea{\begin{eqnarray}}
\def\eea{\end{eqnarray}}
\def\nn{\nonumber}
\def\sss{\scriptscriptstyle}
\def\roughly#1{\mathrel{\raise.3ex\hbox
{$#1$\kern-.75em\lower1ex\hbox{$\sim$}}}}
\def\lsim{\roughly<}

\def\bd{B_d^0}

\def\btod{{\bar b} \to {\bar d}}
\def\btos{{\bar b} \to {\bar s}}
\def\ks{K_{\sss S}}

\def\pewp{P'_{\sss EW}}
\def\pewpc{P'^{\sss C}_{\sss EW}}

\def\pewp{P'_{\sss EW}}

\def\btopik{B \to \pi K}
\def\ApNPqph{{\cal A}^{\prime,q} e^{i \Phi'_q}}
\def\ApNPuph{{\cal A}^{\prime,u} e^{i \Phi'_u}}
\def\ApNPdph{{\cal A}^{\prime,d} e^{i \Phi'_d}}
\def\ApNPCqph{{\cal A}^{\prime {\sss C}, q} e^{i \Phi_q^{\prime C}}}
\def\ApNPCuph{{\cal A}^{\prime {\sss C}, u} e^{i \Phi_u^{\prime C}}}
\def\ApNPCdph{{\cal A}^{\prime {\sss C}, d} e^{i \Phi_d^{\prime C}}}
\def\ApNPcomb{{\cal A}^{\prime, comb} e^{i \Phi'}}

\def\bra#1{\left\langle #1\right|}
\def\ket#1{\left| #1\right\rangle}
\def\ANPq{{\cal A}^q}

\begin{document}

\begin{flushright}
UdeM-GPP-TH-05-137 \\
McGill 05/05 \\
\end{flushright}

\begin{center}
\bigskip
{\Large \bf \boldmath
Polarization States in $B \to \rho K^*$ and New Physics
} \\
\bigskip
\bigskip
{\large Seungwon Baek $^{a,}$\footnote{baek@lps.umontreal.ca},
Alakabha Datta $^{b,}$\footnote{datta@physics.utoronto.ca},
Philippe Hamel $^{a,}$\footnote{philippe.hamel@umontreal.ca}, \\
Oscar F. Hern\'andez $^{c,}$\footnote{oscarh@physics.mcgill.ca;
Permanent Address: Marianopolis College, 3880 Cote-des-Neiges,
Montreal, QC, Canada H3H 1W1 } and David London
$^{a,}$\footnote{london@lps.umontreal.ca}}
\end{center}
\begin{flushleft}
~~~~~~~~~~~$a$: {\it Physique des Particules, Universit\'e
de Montr\'eal,}\\
~~~~~~~~~~~~~~~{\it C.P. 6128, succ. centre-ville, Montr\'eal, QC,
Canada H3C 3J7}\\
~~~~~~~~~~~$b$: {\it Department of Physics, University of Toronto,}\\
~~~~~~~~~~~~~~~{\it 60 St.\ George Street, Toronto, ON, Canada M5S
1A7}\\
~~~~~~~~~~~$c$: {\it Physics Department, McGill University,}\\
~~~~~~~~~~~~~~~{\it 3600 University St., Montr\'eal, Qu\'ebec,
Canada,
H3A 2T8.}\\
\end{flushleft}

\begin{center}
\bigskip (\today)
\vskip0.5cm {\Large Abstract\\} \vskip3truemm
\parbox[t]{\textwidth}{The standard-model explanations of the
anomalously-large transverse polarization fraction $f_{\sss T}$ in
$B\to\phi K^*$ can be tested by measuring the polarizations of the two
decays $B^+ \to \rho^+ K^{*0}$ and $B^+ \to \rho^0 K^{*+}$. For the
scenario in which the transverse polarizations of both $B\to\rho K^*$
decays are predicted to be large, we derive a simple relation between
the $f_{\sss T}$'s of these decays. If this relation is not confirmed
experimentally, this would yield an unambiguous signal for new
physics. The new-physics operators which can account for the
discrepancy in $B\to\pi K$ decays will also contribute to the
polarization states of $B\to\rho K^*$. We compute these contributions
and show that there are only two operators which can simultaneously
account for the present $B \to \pi K$ and $B \to \rho K^*$ data. If
the new physics obeys an approximate U-spin symmetry, the $B\to\phi
K^*$ measurements can also be explained.}
\end{center}

\thispagestyle{empty}
\newpage
\setcounter{page}{1}
\baselineskip=14pt

\section{Introduction}
\label{intro}

One class of $B$ decays which is particularly intriguing involves
processes whose principal contribution comes from $\btos$ penguin
amplitudes. The reason is that there are already several results in
these processes hinting at the presence of physics beyond the standard
model (SM).

First, within the SM, the CP asymmetry in $\bd(t) \to J/\psi \ks$
($\sin 2\beta = 0.725 \pm 0.037$ \cite{sin2beta}) should be
approximately equal to that in penguin-dominated $\btos q{\bar q}$
transitions ($q=u,d,s$). However, on average, these latter
measurements yield a smaller value: $\sin 2\beta = 0.43 \pm 0.07$
\cite{sin2betapeng}. 

Second, within the SM, one expects no triple-product asymmetries in $B
\to \phi K^*$ \cite{BVVTP}. However both BaBar and Belle have measured
such effects, albeit at low statistical significance
\cite{phiKstarTP}.

Third, the latest data on $B\to\pi K$ branching ratios and CP
asymmetries \cite{BKpiexp} appear to be inconsistent with a SM fit
\cite{BKpidecays1, BKpidecays}. The model-independent analysis in
Ref.~\cite{BKpidecays} has shown that the data can be accommodated
with a new-physics (NP) operator in the electroweak penguin sector.

A fourth possible hint of NP occurs in $B\to V_1 V_2$ decays, where
the $V_i$ are light vector mesons. In such decays the final-state
particles can be found with transverse or longitudinal
polarization. SM factorizable amplitudes, which are expected to
dominate in the heavy $b$-quark limit, result in a dominant
longitudinal polarization, with the transversely-polarized amplitudes
suppressed by $m_{\sss V}/m_{\sss B}$. While this is realized for
$B\to\rho\rho$ decays, which receive $\btod$ penguin contributions, in
$B \to \phi K^*$ decays it is found that the transverse fraction
$f_{\sss T}$ is about equal to the longitudinal fraction $f_{\sss L}$
\cite{pdg,phiKstarexp}. Competing NP \cite{phiKstarNP,DasYang}, and SM
\cite{colangelo,Kagan,phiKstarSMHou} explanations have been
proposed. $B\to\rho K^*$ decays may offer a way to resolve this
discrepancy.

In this paper we will be mainly focussing our attention on the third
and fourth points above. In the decay $B\to \rho K^*$, unlike
$B\to\phi K^*$, there are two states, distinguished by the charge of
the $\rho$ meson: $\rho^+$ or $\rho^0$. Here, the final-state
particles are also vector mesons, so that one can measure their
polarization states. Now, the polarization states of $B\to \rho K^*$
can be related to those in $B\to\phi K^*$. For this latter decay, it
is not clear whether the large observed value of $f_{\sss T}/f_{\sss
L}$ is accommodated by the SM or best explained with NP.  However one
can distinguish between a SM and NP explanation by {\it comparing} the
two charge states. In particular, we show that if one of the SM
scenarios proposed in Refs.~\cite{colangelo,Kagan} does explain the
large $B\to\phi K^*$ transverse polarization, then the transverse
fractions of the two charge states in $B\to \rho K^*$ should satisfy
$f_{\sss T}^+/f_{\sss T}^0 \simeq 2 ({\rm BR}^0/{\rm BR}^+)$.
Alternatively, if the SM scenario for the $B\to\phi K^*$ modes in
Ref.~\cite{phiKstarSMHou} is correct, then the $f_{\sss L}$ fraction
of both charged $B\to \rho K^*$ decays should be greater than 90\%. If
neither of these two results is observed then non-SM physics is
involved in the decays. We derive and discuss these prediction in
Sec.~\ref{smpredictions}.

The decay $B\to \rho K^*$ is described at the quark level by $\btos
q{\bar q}$ ($q=u,d$). This is the same quark-level decay that
contributes to $B\to\pi K$. If there is NP in these latter decays, it
will affect $B\to \rho K^*$. Thus, given a $B\to\pi K$ NP scenario, we
can examine its effects on the $B\to \rho K^*$ polarizations. We
review the data on $B\to\pi K$ decays, as well as the size of NP
operators which can account for it, in Sec.~\ref{btopik}. 

Sec.~\ref{btorhok} contains the calculation of the contribution of
these NP operators to the polarization states of charged $B\to\rho
K^*$ decays. Under some simplifying assumptions we show that only NP
operators of the form ${\bar b} \gamma_{\sss R} s \, {\bar d}
\gamma_{\sss R} d$ or ${\bar b} \gamma_{\sss L} s \, {\bar d}
\gamma_{\sss L} d$ can explain both the $\pi K$ and $\rho K^*$
data. We then discuss ways of testing this scenario.

Finally, in Sec.~\ref{btophik} we examine the consequences of the NP
scenario for $B\to\phi K^*$ decays. We show that if the NP respects an
approximate U-spin symmetry, it can simultaneously account for the
$\pi K$, $\rho K^*$ and $\phi K^*$ data.  We conclude in
Sec.~\ref{conclusions}.

\section{$B\to\rho K^*$: Standard Model Predictions}
\label{smpredictions}

Before examining the contributions of new physics to the polarization
states in $B\to\rho K^*$ decays, it is first necessary to understand
the SM predictions for these states.

In the following, we denote $A_0$ as the longitudinal polarization
amplitude for a decay, and $A_{++}$ and $A_{--}$ as the amplitudes
with both vector mesons in the right-handed or left-handed helicity
state, respectively. The transverse amplitudes are then
$A_\|=(A_{++}+A_{--})/\sqrt2$ and $A_\perp=(A_{++}-A_{--})/\sqrt2$,
while the total amplitude squared is $ |A_{\rm total}|^2
=|A_0|^2+|A_\||^2+|A_\perp|^2$. The individual polarization fractions
are
\beq 
f_{\sss L} = {|A_0|^2 \over |A_{\rm total}|^2} ~~,~~~~ f_\| = {|A_\||^2
\over |A_{\rm total}|^2} ~~,~~~~ f_\perp = {|A_\perp|^2 \over |A_{\rm
total}|^2}.
\label{fractions} 
\eeq
For a given decay, the branching ratio is related to the polarization
amplitudes by
\beq 
{\rm BR} = (|A_0|^2 + |A_\||^2 + |A_\perp|^2) \, {\rm PS} /
\Gamma_{\rm total} ~,
\label{BR}
\eeq
where ${\rm PS}$ is a phase-space factor, and $\Gamma_{\rm total}$
is the total decay width .

It is useful to express the amplitudes for the various decays in terms
of diagrams \cite{GHLR}. These include a ``tree'' amplitude $T'$, a
``color-suppressed'' amplitude $C'$, a gluonic ``penguin'' amplitude
$P'$, a color-favored electroweak penguin (EWP) amplitude $\pewp$ and
a color-suppressed EWP amplitude $\pewpc$. Other diagrams are
higher-order in $1/m_{\sss B}$ and are expected to be smaller. They
will be neglected in our calculations. Here the prime on the amplitude
stands for a strangeness-changing decay.

The diagram $P'$ in fact includes three pieces, corresponding to the
internal quarks $u$, $c$ and $t$.
\beq
P' = V_{ub}^* V_{us} \, P'_u + V_{cb}^* V_{cs} \, P'_c + V_{tb}^*
V_{ts} \, P'_t = V_{ub}^* V_{us} \, P'_{ut} + V_{cb}^* V_{cs} \,
P'_{ct} ~.
\label{penguin}
\eeq
Here, $P'_{qt}=P'_q - P'_t$ ($q=u,c$). On the right-hand side, the
unitarity of the Cabibbo-Kobayashi-Maskawa (CKM) matrix has been used
to reduce the number of terms. Since $|V_{ub}^* V_{us}| \ll |V_{cb}^*
V_{cs}|$, only the last term above is important; the first piece can
be neglected. In addition, $\pewpc$ and $C'$ are expected to be
smaller than $\pewp$ and $T'$ \cite{GHLR}, and will also be neglected
in our calculations. Our amplitudes will therefore be expressed in
terms of the diagrams $P'_{ct}$, $T'$ and $\pewp$.

Furthermore, it has been shown that, to a good approximation, the EWP
amplitude $\pewp$ can be related to $T'$ \cite{EWPs}:
\beq
\pewp \simeq {3\over 4} \left[ {c_9 + c_{10} \over c_1 + c_2} + {c_9 -
c_{10} \over c_1 - c_2} \right] {1 \over \lambda^2} {\sin(\beta
+\gamma) \over \sin\beta} \, T' \equiv - Z \, T' \simeq -0.65 \, T' ~,
\label{pewprel}
\eeq
where $\lambda = 0.22$ is the Cabibbo angle, $\beta$ and $\gamma$ are
CP phases (the phase information in the CKM quark mixing matrix is
conventionally parametrized in terms of the unitarity triangle, in
which the interior (CP-violating) angles are known as $\alpha$,
$\beta$ and $\gamma$ \cite{pdg}), and the $c_i$ are (known) Wilson
coefficients \cite{BuraseffH}.

We begin with a study of $B\to\phi K^*$. This is a pure penguin decay
whose amplitude can be written
\beq
A(B \to \phi K^*) \simeq P'_{ct}- {1 \over 3}\pewp-{1 \over 3} \pewpc `.
\eeq
The penguin operator $P'_{ct}$ has $(V-A) \times (V-A)$ and $(V-A)
\times (V+A)$ pieces while the EWP's have mainly $(V-A) \times (V-A)$
structure. For operators with $(V-A) \times (V \mp A)$ structure, a
single spin flip is required to produce the $A_{--}$ amplitude and a
double spin flip for the $A_{++}$ amplitude. Each spin flip leads to a
$1/m_{\sss B}$ suppression, causing the amplitudes $A_\perp$ and
$A_\|$ to be $1/m_{\sss B}$ suppressed. Thus, the SM operators
naturally contribute mainly to the longitudinal polarization in $B \to
\phi K^*$; their transverse polarization contribution is down by at
least $ O(1/m_{\sss B}^2)$ relative to the total decay amplitude. The
SM predictions for this decay can then be written as
\beq
f_{\sss L} = 1- O(1/m_{\sss B}^2) ~~,~~~~ f_{\sss T} = O(1/m_{\sss B}^2)
~~,~~~~ { f_{\perp} \over f_\|} = 1 + O(1/m_{\sss B}) ~.
\label{smprediction}
\eeq
The large transverse polarization observed in $B\to\phi K^*$ is then a
puzzle for the SM.

However, there may be certain sources of large transverse polarization
within the SM.  Rescattering effects from tree-level $\btos c{\bar c}$
operators have been identified as a possible source of large
transverse polarization \cite{colangelo}. In Eq.~(\ref{penguin}) this
effect is represented by $P'_c$ and is contained in $P'_{ct}$. The
claim here is then that rescattering effects from $P'_c$ can enhance
one or both of the transverse amplitudes associated with
$P'_{ct}$. 

Another possible source for the enhancement of the transverse
amplitudes is associated with $P'_{ct}$ though annihilation topologies
\cite{Kagan}. The dominant contribution comes from the $(S-P) \times
(S+P)$ operators in the effective Hamiltonian, produced by performing
a Fierz transformation on the $(V-A) \times (V+A)$ piece of $P'_{ct}$.
Even though formally suppressed in the heavy $m_b$ limit, these
contributions can produce an $O(1)$ effect on the transverse
polarization amplitudes due to large coefficients.

Finally, a third SM explanation for the large transverse polarization
in $B \to \phi K^*$ is proposed in Ref.~\cite{phiKstarSMHou}. Here,
the transverse amplitudes are enhanced because the gluon from the
$\btos g$ transition hadronizes directly into the $\phi$, with the
exchange of additional gluons to take care of color factors.

We therefore see that it may be possible to account for the large
transverse polarization in $B\to \phi K^*$ through SM effects.
Fortunately, it is possible to test these explanations through the
measurement of the transverse polarization in $B\to\rho K^*$
decays. The key point here is that, in contrast to $B\to\phi K^*$,
there are {\it two} decays, $B \to \rho^+ K^*$ and $B \to \rho^0
K^*$. It is the measurement of the polarization states of both decays
which allows us to distinguish the various explanations of the
$B\to\phi K^*$ data. In the following, we concentrate on charged $B$
decays; the discussion is similar when neutral $B$'s are involved. We
use the indices `$+$' and `$0$' to indicate the decays $B^+ \to \rho^+
K^{*0}$ and $B^+ \to \rho^0 K^{*+}$, respectively.

In the SM, neglecting the small amplitudes, the two $B^+\to \rho K^*$
amplitudes are given by
\bea
A(B^+ \to \rho^+ K^{*0}) \equiv A^+ & = & P'_{ct} ~, \nn\\
\sqrt{2} A(B^+ \to \rho^0 K^{*+}) \equiv \sqrt{2} A^0 & = & - P'_{ct}
- T' \, e^{i\gamma} - \pewp ~.
\label{finalamp}
\eea
We have explicitly written the dependence on the weak phase $\gamma$,
but the amplitudes contain strong phases. These amplitudes allow us to
test the SM explanations of the large transverse polarization in
$B\to\phi K^*$ by comparing the two $B\to\rho K^*$ decays. In
particular, we calculate the transverse polarization pieces of
\beq
{ \left\vert A^+ \right\vert^2 - 2 \left\vert A^0 \right\vert^2 \over
\left\vert A^+ \right\vert^2} ~.
\label{SMreln}
\eeq

Consider first Ref.~\cite{colangelo}, which invokes rescattering from
tree-level $\btos c{\bar c}$ operators, so that $P'_{ct}$ is
affected. The rescattering represented by $P'_u$ ($\btos u{\bar u}$
operators) is small because of CKM suppression, so that the amplitudes
$T'$ and $\pewp$ are essentially unaffected.  Ref.~\cite{Kagan} is
similar.  Here, large annihilation effects modify $P'_{ct}$; the
amplitudes $T'$ and $\pewp$ remain effectively unchanged. In both
cases, the change in $P'_{ct}$ persists in $B\to\rho K^*$ decays, so
that a large transverse polarization in these processes is expected.
Since both decays are dominated by $P'_{ct}$, to leading order the
numerator of Eq.~\ref{SMreln} vanishes, and it is predicted that
\beq
f_{\sss T}^+ = 2 f_{\sss T}^0 \left( {{\rm BR}^0 \over {\rm BR}^+}
\right) ~.
\label{fTreln}
\eeq
The systematic error in this relation comes from the contribution of
$T'$ to the transverse polarization, which is suppressed by $m_{\sss
V}/m_{\sss B}$:
\beq
{\rm sys} = O \left( 2 {T' \over P'_{ct}} {m_{\sss V}
\over m_{\sss B}} \right) \sim 10\% ~.
\eeq
We repeat that this systematic error holds only for the case in which
the transverse polarization in both $B\to\rho K^*$ decays is large. If
it is small, then the systematic error is correspondingly larger.

In the third SM explanation \cite{phiKstarSMHou}, the transverse
amplitude in $B \to \phi K^*$ is enhanced due to direct gluon
hadronization into the $\phi$. Since the gluon has isospin zero, there
should be no effect on $B\to\rho K^*$. Thus, in this model the usual
SM arguments apply to both decay modes, giving a $f_{\sss T}$ that is
suppressed by $(m_{\sss V}/m_{\sss B})^2$.

These qualitative arguments can be made quantitative. We note that the
amplitudes given in Eq.~(\ref{finalamp}) apply to the longitudinal and
transverse polarizations individually. Thus, the transverse pieces
($T= \perp, \|$) of the two amplitudes are related as
\beq
\sqrt{2} (A^0)_{\sss T} = - (A^+)_{\sss T} \left [ 1+ x_{\sss T} e^{ i
\Delta_{\sss T}} \right] ~,
\label{tamp}
\eeq
with
\beq
x_{\sss T} e^{ i \Delta_{\sss T}} \equiv {T'_{\sss T} \, e^{i\gamma} +
P'_{\sss EW,T} \over P'_{\sss T}} = {T'_{\sss T}( \, e^{i\gamma} - Z)
\over P'_{\sss T}} ~.
\label{amperror}
\eeq

Now, because QCD respects isospin symmetry, the phase factors in
Eq.~(\ref{BR}) for both $B^+ \to \rho^+ K^{*0}$ and $B^+ \to \rho^0
K^{*+}$ are equal to within a few percent. Thus, a prediction of the
SM using Eq.~(\ref{tamp}) is that the transverse polarizations in both
charge states of $B\to\rho K^*$ should be related. At leading order,
$\sqrt{2} (A^0)_{\sss T} = - (A^+)_{\sss T}$, so that
\beq
E_{\sss T} = {f_{\sss T}^+ {\rm BR}^+ - 2 f_{\sss T}^0 {\rm BR}^0
 \over f_{\sss T}^+ {\rm BR}^+ } \approx 0 ~.
\label{ftrelation}
\eeq

The systematic error in this relation, $\Delta E_{\sss T}$, can be
estimated by keeping terms linear in $x_{\sss T}$:
\beq
\Delta E_{\sss T} \approx -2x_{\sss T} \cos {\Delta_{\sss T}} ~~,~~~~
x_{\sss T} \approx {{|T'_{\sss T}|} \over {|P'_{\sss L}|}}(1+Z^2-2Z\cos
\gamma)^{1/2} { \sqrt{ f_{\sss L}^+ \over f_{\sss T}^+}} ~,
\label{error}
\eeq
where $P'_{\sss L}$ is the longitudinal contribution from $P'$.  Using
$|T'_{\sss T}| \sim (m_{\sss K^*} / m_{\sss B}) |T'_{\sss L}|$ and
taking $\left\vert T'_{\sss L} / P'_{\sss L} \right\vert \sim 0.4$
\cite{BVVTP}, we find
\beq
|\Delta E_{\sss T}| \lsim 10 \% { \sqrt{ f_{\sss L}^+ \over f_{\sss
 T}^+}} ~.
\label{errorestimate}
\eeq
{}From this expression, we see that a large value of $\sqrt{ f_{\sss
T}^+ / f_{\sss L}^+}$ would result in a smaller systematic error in
Eq.~(\ref{ftrelation}). Thus, this relation is most useful if a large
transverse polarization is observed in the $\rho^+ K^*$ mode.

Relations involving the longitudinal polarizations will have errors of
the order of $x_{\sss L} \sim (m_{\sss B} / m_{K^*})x_{\sss T}$, which
can be significant. Additional measurements, such as direct CP
asymmetries and triple-product asymmetries in both $\rho K^*$ modes
would provide important constraints on the various amplitudes and
their phases, thereby providing strong tests of the SM.

\begin{table}
\caption{Branching ratios and polarization fractions for the two $B^+
\to \rho K^*$ decays. Data comes from Ref.~\cite{BrhoKstarBRsfL};
averages are taken from Ref.~\cite{HFAG}.}
\begin{center}
\begin{tabular}{c|c|c|c}
\hline 
\hline 
& & $B^+ \to \rho^+ K^{*0}$ & $B^+ \to \rho^0 K^{*+}$ \\
\hline 
  $BR[10^{-6}]$ & Belle & $8.9 \pm 1.7 \pm 1.2$ & \\
  & BaBar & $17.0 \pm 2.9 {}^{+2.0}_{-2.8}$ & $10.6 {}^{+3.0}_{-2.6} \pm 2.4$ \\
  & average & $10.6 \pm 1.9$ & $10.6 {}^{+3.8}_{-3.5}$ \\
\hline 
  $f_{\sss L}$ & Belle & $0.43 \pm 0.11 {}^{+0.05}_{-0.02}$ & \\
  & BaBar & $0.79 \pm 0.08 \pm 0.04$ & $0.96 {}^{+0.04}_{-0.15} \pm 0.04$ \\
  & average & $0.66 \pm 0.07$ & $0.96 {}^{+0.06}_{-0.15}$ \\
\hline
\end{tabular}
\end{center}
\label{BrhoKstardata}
\end{table}

The above SM predictions can now be compared with the present
$B\to\rho K^*$ data, shown in Table~\ref{BrhoKstardata}. Using the
central values, and using the SM relation $A_{\perp} \approx A_{\|}$,
we find $E_{\perp} \approx E_{\|} \approx 77\%$. This is very far from
the expected value of zero, so that one might be tempted to claim the
presence of new physics. However, even though the systematic error
$\Delta E_{\perp} \approx \Delta E_{\|}$ is relatively small, $\sim
20\%$, the statistical error is enormous, $\pm 129 \%$. Thus, the
errors are still much too large to claim any discrepancy with the
SM. However, this does demonstrate the importance of more precise
measurements of the polarizations in $B\to\rho K^{*}$ decays.

While the predictions of Refs.~\cite{colangelo,Kagan} are not
invalidated, the same is not true for Ref.~\cite{phiKstarSMHou}. In
this scenario, the $f_{\sss L}$ fraction of both charged $B\to \rho
K^*$ decays is predicted to be greater than 90\%. However, the data in
Table~\ref{BrhoKstardata} show that this clearly does not hold for
$B^+ \to \rho^+ K^{*0}$, ruling out this SM explanation at the
$3.5\sigma$ level.

Finally, we note that in the pQCD approach, even with annihilation and
nonfactorizable effects, the large transverse polarization in
$B\to\phi K^*$ cannot be explained \cite{pqcd1}. In Ref.~\cite{pqcd2},
it is argued that one of the $B \to K^*$ form factors must be reduced
to explain the $B\to\phi K^*$ polarization. It is not clear whether
this can be done, but the prediction of this scenario is then that the
$B\to\phi K^*$ longitudinal polarization is smaller than that of both
the $B^+\to\rho^+ K^*$ and $B^+\to\rho^0 K^{+*}$ modes. The careful
measurement of the polarization fractions in the $B \to \rho K*$ modes
will test this scenario.

\section{$B\to\pi K$ Decays}
\label{btopik}

There are four $B\to\pi K$ decays. In the SM, neglecting small
diagrams as usual, their amplitudes are given by
\bea
\label{SMamps}
A(B^+ \to \pi^+K^0) \equiv A^{+0} & = & P'_{ct} ~, \nn\\
\sqrt{2} A(B^+ \to \pi^0K^+) \equiv \sqrt{2} A^{0+} & = & -T'
e^{i\gamma} -P'_{ct} - \pewp , \nn\\
A(B^0\to \pi^-K^+) \equiv A^{-+} & = & -T' e^{i\gamma} -P'_{ct} ~, \nn\\
\sqrt{2} A(B^0\to \pi^0K^0) \equiv \sqrt{2}A^{00} & = & P'_{ct} - \pewp ~,
\eea
(Isospin implies the relation $A^{+0}+ \sqrt{2}
A^{0+}=A^{-+}+\sqrt{2}A^{00}$.) It is difficult to explain the present
data (branching ratios, CP asymmetries) using only this
parametrization \cite{BKpidecays}.

We therefore consider the addition of new $\btos q{\bar q}$
($q=u,d$) operators. One can show that the strong phase of any NP
operator is much smaller than that of the SM \cite{DLNP}. In this
case, for a given type of transition, all NP matrix elements can
now be combined into a single effective NP amplitude, with a single
weak phase:
\beq 
\sum \bra{\pi K} {\cal O}_{\sss NP}^{q} \ket{B} = \ANPq e^{i
\Phi_q} ~, 
\eeq
in which the symbols ${\cal A}$ and $\Phi$ denote the NP amplitudes
and weak phases, respectively. In $B\to\pi K$ decays, there are four
classes of NP operators, differing in their color structure: ${\bar
b}_\alpha \Gamma_i s_\alpha \, {\bar q}_\beta \Gamma_j q_\beta$ and
${\bar b}_\alpha \Gamma_i s_\beta \, {\bar q}_\beta \Gamma_j q_\alpha$
($q=u,d$). The matrix elements of these operators can be combined into
single NP amplitudes, denoted $\ApNPqph$ and $\ApNPCqph$, respectively
\cite{BNPmethods}. Each of these contributes differently to the
various $\btopik$ decays. (Note that, despite the color-suppressed
index $C$, the matrix elements $\ApNPCqph$ are not necessarily smaller
than the $\ApNPqph$.)

In the presence of these NP matrix elements, the $\btopik$
amplitudes take the form \cite{BKpidecays,BNPmethods}:
\bea \label{BpiKNPamps}
A^{+0} &\!\!=\!\!& P'_{ct} + \ApNPCdph ~, \nn\\
\sqrt{2} A^{0+} &\!\!=\!\!& - P'_{ct} - T' \, e^{i\gamma} - \pewp
+ \ApNPcomb - \ApNPCuph ~, \nn\\
A^{-+} &\!\!=\!\!& - P'_{ct} - T' \, e^{i\gamma} - \ApNPCuph ~, \nn\\
\sqrt{2} A^{00} &\!\!=\!\!& P'_{ct} - \pewp + \ApNPcomb + \ApNPCdph ~,
\eea
where $\ApNPcomb \equiv - \ApNPuph + \ApNPdph$.

Even taking into account the fact that $\pewp$ and $T'$ are related
\cite{EWPs}, there are too many theoretical parameters to perform a
fit. For this reason, the authors of Ref.~\cite{BKpidecays} assumed
that a single NP amplitude dominates. They considered four
possibilities: (i) only ${\cal A}^{\prime, comb} \ne 0$, (ii) only
${\cal A}^{\prime {\sss C}, u} \ne 0$, (iii) only ${\cal A}^{\prime
{\sss C}, d} \ne 0$, (iv) $\ApNPCuph = \ApNPCdph$, ${\cal A}^{\prime,
comb} = 0$ (isospin-conserving NP).  Of these, only choice (i) gave a
good fit; the others produced poor or very poor fits\footnote{Note
that the poor fit gave a discrepancy of only about $2\sigma$ with the
SM, so that, strictly speaking, it cannot be ruled out. However, in
what follows, we concentrate on the good fit.}. The good fit found
best-fit values of $| {\cal A}^{\prime, comb} / P' | = 0.36$ and
$|T'/P'| = 0.22$. Thus, the NP parameter was found to be larger than
the tree amplitude, with $| {\cal A}^{\prime, comb} / T' | = 1.64$.

In what follows, we assume that NP of type (i) is present in $B\to\pi
K$ decays. This same NP will affect $B\to\rho K^*$ decays. In order to
calculate the effect on the $B\to\rho K^*$ polarization states, we
must assume a particular form for ${\cal A}^{\prime, comb}$. There are
many NP operators which can contribute to ${\cal A}^{\prime,
comb}$. They are
\beq 
\label{NPoperators} 
{4 G_{\sss F} \over \sqrt{2}} \sum_{\sss A,B = L,R} \left\{ f_q^{\sss
AB} \, {\bar b} \gamma_{\sss A} s \, {\bar q} \gamma_{\sss B} q +
g_q^{\sss AB} \, {\bar b} \gamma^\mu \gamma_{\sss A} s \, {\bar q}
\gamma_\mu \gamma_{\sss B} q \right\} ~.
\eeq
There are a total of 16 contributing operators ($A,B = L,R$, $q=u,d$);
tensor operators do not contribute to $B\to\pi K$. For simplicity, we
assume that a single operator contributes to ${\cal A}^{\prime,
comb}$, and we analyze their effects one by one.

Note that all operators contribute directly to $\pi K$ final states
involving a $\pi^0$. They can also contribute to states involving a
$\pi^+$ if one performs Fierz transformations of the fermions and
colors. However, the effects on $\pi^+ K^0$ are all suppressed by at
least $1/N_c$, so that the contributions to $\pi^0 K^+$ are
larger. This is approximately consistent with the hypothesis of
including only ${\cal A}^{\prime, comb}$.

We begin by considering the operators whose coefficients are
$f_q^{\sss AB}$ [Eq.~(\ref{NPoperators})]. Using $| {\cal A}^{\prime,
comb} / T' | = 1.64$ and
\beq T' = {G_{\sss F} \over \sqrt{2}} \, V_{ub}^* V_{us} \left(
c_1 + {c_2 \over N_c} \right) \bra{\pi^0 K^+} {\bar u} \gamma^\mu
(1 - \gamma_5) s \, {\bar b} \gamma_\mu (1 - \gamma_5) u \ket{B^+}
~, \eeq
where $c_1 = 1.081$ and $c_2 = -0.190$ are the Wilson coefficients
characterizing $T'$ \cite{BuraseffH}, we can estimate the size of the
NP coefficients. To do this, we use naive factorization. This is
reasonable since we are interested only in estimates. More accurate
calculations can use a more precise formalism, e.g.\
Ref.~\cite{beneke}.

We then have:
\beq \left\vert { 4 f_q^{\sss AB} \bra{\pi^0} {\bar q}
\gamma_{\sss B} q \ket{0} \bra{K^+} {\bar b} \gamma_{\sss A} s
\ket{B^+} \over V_{ub}^* V_{us} \bra{\pi^0} {\bar b} \gamma_\mu (1
- \gamma_5) u \ket{B^+} \bra{K^+} {\bar u} \gamma^\mu (1 -
\gamma_5) s \ket{0} } \right\vert = 1.64 ~. \eeq
Using the matrix elements given in the Appendix, we find
\beq \left\vert f_q^{\sss AB} \right\vert = { f_{\sss K} (m_{\sss
B}^2 - m_\pi^2) F_0^\pi / \sqrt{2} \over [ ( m_{\sss B}^2 -
m_{\sss K}^2) / (m_b - m_s) ] F_0^{\sss K} ( m_\pi^2 / 2 m_q )
f_\pi / \sqrt{2} } \, 1.64 \left( c_1 + {c_2 \over N_c} \right)
\left\vert V_{ub}^* V_{us} \right\vert ~. \eeq
We take $(f_{\sss K}/f_\pi) (F_0^\pi/F_0^{\sss K}) \sim 1$,
$|V_{ub}^* V_{us}/V_{tb}^* V_{ts}| = 1/48$ and $c_1 + c_2/N_c =
1.018$. Taking the masses from the Particle Data Group \cite{pdg},
we find
\bea \left\vert f_d^{\sss LL} \right\vert = \left\vert f_d^{\sss
RR} \right\vert = \left\vert f_d^{\sss LR} \right\vert =
\left\vert f_d^{\sss RL} \right\vert & = & \cases{ 0.069 |V_{tb}^*
V_{ts}| & $m_d = 4$ MeV, \cr 0.138 |V_{tb}^* V_{ts}| &
$m_d = 8$ MeV, \cr} \nn\\
\left\vert f_u^{\sss LL} \right\vert = \left\vert f_u^{\sss RR}
\right\vert = \left\vert f_u^{\sss LR} \right\vert = \left\vert
f_u^{\sss RL} \right\vert & = & \cases{ 0.026 |V_{tb}^* V_{ts}| &
$m_u = 1.5$ MeV, \cr 0.069 |V_{tb}^* V_{ts}| & $m_u = 4$ MeV. \cr}
\label{coeffvalues} \eea

The operators associated with the parameters $g_q^{\sss AB}$
[Eq.~(\ref{NPoperators})] can be analyzed similarly. The sizes of the
NP coefficients are
\beq \left\vert g_q^{\sss LL} \right\vert = \left\vert g_q^{\sss
RR} \right\vert = \left\vert g_q^{\sss LR} \right\vert =
\left\vert g_q^{\sss RL} \right\vert = 0.035 |V_{tb}^* V_{ts}|
~,~~ q = u,d ~. \eeq

We remind the reader that we have assumed that a single NP operator
contributes to ${\cal A}^{\prime, comb}$. For each operator, we have
calculated the size of the coefficient which reproduces the $B \to \pi
K$ data. These same operators will affect the $B\to\rho K^*$
polarization states. We compute these effects in the next section.

\section{$B\to\rho K^*$: New-Physics Contributions}
\label{btorhok}

If there is new physics in $B\to\pi K$ decays, it is of the form
$\btos q{\bar q}$ ($q=u,d$), and will, in general, contribute to
$B\to\rho K^*$ decays. In this section, we proceed as above, and
calculate the effect on $B\to\rho K^*$ of each of the operators in
Eq.~(\ref{NPoperators}).

We begin with some general statements. The amplitude for an arbitrary
$B\to V_1 V_2$ decay can be written as (for example, see
Ref.~\cite{BVVTP})
\beq 
{\cal M} = a \, \epsilon^*_1 \cdot \epsilon^*_2 + {b \over m_{\sss
B}^2} \left( \epsilon^*_1 \cdot p_2 \right) \left( \epsilon^*_2 \cdot
p_1 \right) - 2 i {c \over m_{\sss B}^2} \epsilon_{\mu\nu\alpha\beta}
\, p_1^\mu p_2^\nu \epsilon_1^\alpha \epsilon_2^\beta ~,
\label{BVVamp} 
\eeq
with
\beq
A_\| = \sqrt{2} a ~,~~~ A_0 = -a x - {m_1 m_2 \over m_{\sss B}^2} b
(x^2 - 1) ~,~~~ A_\perp = 2\sqrt{2} \, {m_1 m_2 \over m_{\sss B}^2} c 
\sqrt{x^2 - 1} ~,
\label{Aidefs}
\eeq
where $x = p_1 \cdot p_2 / (m_1 m_2)$. Here we are considering $B \to
V_1 V_2$ decays in which the final vector mesons are light: $m_{1,2}
\ll m_{\sss B}$. Neglecting terms of $O(m_{1,2}^2/m_{\sss B}^2)$, we
can then approximate $E_1 \sim E_2 \sim |\vec{k}|= E = m_{\sss
B}/2$. Then, using Eq.~(\ref{Aidefs}), we have for the various linear
polarization amplitudes
\beq
A_0 \approx -(2a+b) \frac{E^2}{m_1m_2} ~~,~~~~ A_\| \approx \sqrt{2}a
~~,~~~~ A_\perp \approx \sqrt{2}c ~.
\label{smallmass}
\eeq
The procedure for computing the SM or NP contributions to polarization
amplitudes is then clear: we first express the amplitude for a
particular $B\to V_1 V_2$ decay as in Eq.~(\ref{BVVamp}) and then use
the above relations to obtain $A_0$, $A_\|$ and $A_\perp$. For the SM,
in which all operators have $(V-A) \times (V \mp A)$ structure, one
can show that $2a+b \sim m_{\sss V}/m_{\sss B}$, so that the
polarization fractions are predicted to be as in
Eq.~(\ref{smpredictions}). (We will see this explicitly below for $B^+
\to \rho^+ K^{*0}$.)

The present data is consistent with the SM expectations for $B^+ \to
\rho^0 K^{*+}$, but suggests that there may be new physics in $B^+ \to
\rho^+ K^{*0}$. For this reason, we concentrate on this latter decay
in what follows.

Using factorization, the SM amplitude for the decay $B^+ \to \rho^+
K^{*0}$ is given by
\beq 
A[B^+ \to \rho^+ K^{*0} ] = \frac{G_F}{\sqrt{2}} [X_{\rho}
P_{K^*}^{\rho}],
\label{Brhoplus}
\eeq
with
\bea
X_{\rho}  & = &  - \sum_{q=u,c,t} V_{qb}
V_{qs}^* \left( a_4^q - \frac{1}{2} a_{10}^q \right) ~, \nn\\
P_{K^*}^{\rho} & = & m_{K^*}g_{K^*}\varepsilon^{*\mu}_{K^*}
\bra{\rho^+} \bar{d} \gamma_{\mu}(1-\gamma_5)b \ket{B^+} ~, \
\label{Brhoplusmatrix}
\eea
The above amplitude depends on combinations of Wilson coefficients,
$a_i$, where $a_i= c_i+{c_{i+1}/ N_c}$ for $i$ odd and $a_i=
c_i+{c_{i-1}/ N_c}$ for $i$ even. The terms described by the various
$a_i$'s can be associated with the different decay topologies
introduced earlier. The term proportional to $a_4$ is the
color-allowed penguin amplitude, $P'$. The dominant electroweak
penguin $\pewp$ is represented by term proportional to $a_9$, $\pewpc$
is $a_{10}$, and $a_7$ and $a_8$ are additional small EWP amplitudes.
(If there were terms proportional to $a_1$ and $a_2$, they would
represent the color-allowed and color-suppressed tree amplitudes $T'$
and $C'$, respectively.) The values of the Wilson coefficients can be
found in Ref~\cite{BuraseffH}.

Using the matrix elements found in the Appendix, this amplitude can be
put in the form of Eq.~(\ref{BVVamp}). The polarization amplitudes are
then given by 
\bea 
A_{0} & \approx & { G_{\sss F} \over \sqrt{2}} 2 m_{\sss B} m_{K^*}
g_{K^*} X_{\rho} \left[ \left( A_1^{\rho} - A_2^{\rho} \right) +
\frac{m_{\rho}}{m_{\sss B}} \left( A_1^{\rho} + A_2^{\rho} \right)
\right] \frac{m_{\sss B}^2}{4 m_\rho m_{K^*}} ~, \nn\\
A_\| & \approx & - { G_{\sss F} \over \sqrt{2}} \sqrt{2} m_{\sss B}
\left[ m_{K^*} g_{K^*} \left( 1 + {m_{\rho}\over m_{\sss B}} \right)
A_1^{\rho} (m_{K^*}^2) X \right] ~, \nn\\
A_\perp & \approx & - { G_{\sss F} \over \sqrt{2}} \sqrt{2} m_{\sss B}
\left[ m_{K^*} g_{K^*} \left( 1 - {m_{\rho}\over m_{\sss B}} \right)
V^{\rho}(m_{K^*}^2) X \right]~.
\label{rkokstarplus}
\eea
In the large-energy limit, the form factors are related
\cite{formfactors}:
\beq
A_1=A_2 +O(m_{\sss V}/m_{\sss B}) ~~,~~~~ V=A_1+O(m_{\sss V}/m_{\sss
B}) ~.
\label{leet}
\eeq
We therefore find the same suppression of the $A_{\|,\perp}$
amplitudes relative to $A_0$ as was found from helicity arguments
[Eq.~(\ref{smpredictions})].  We therefore see that the SM naturally
predicts the longitudinal polarization for the decay $B^+ \to \rho^+
K^{*0}$ to be enhanced by $O(m_{\sss B}/m_{\sss V})$.

In our simplified approach we will assume the form-factor relations
above and ignore possible power-suppressed and $\alpha_s$ corrections
to them. We then have
\bea
 A_{1}^{\rho} & \approx & \zeta_\perp( 1-{m_{\rho}\over m_{\sss B}}) \nonumber\\
 A_{2}^{\rho} & \approx &
\zeta_\perp( 1+{m_{\rho}\over m_{\sss B}}) - 
{2 m_\rho \over m_{\sss B}} \zeta_\| \nonumber\\
V_{1}^{\rho}  & \approx & \zeta_\perp( 1+{m_{\rho}\over m_{\sss B}}).\
\eea
Choosing $\zeta_\perp \approx \zeta_\|$ gives $A_{1}^{\rho} \approx
A_{2}^{\rho}$, and hence the SM prediction is that
\bea 
A_0^{\sss SM} &\approx & { G_{\sss F} \over \sqrt{2}}
g_{\sss K^*}  m_{\sss B}^2  \cdot \, X \zeta_\| ~,\nonumber\\
A_\|^{\sss SM} &\approx&  -G_{\sss F}
g_{\sss K^*} m_{K^*}  m_{\sss B} \cdot \, X \zeta_\|,   \nonumber\\
A_\perp^{\sss SM} &\approx&  
-G_{\sss F}
g_{\sss K^*} m_{K^*}  m_{\sss B} \cdot \, X \zeta_\| ~,
\label{SMcont} 
\eea
where $X \simeq - a_4^t |V_{tb}^* V_{ts}| = 0.035 |V_{tb}^* V_{ts}|$.

We now turn to the new-physics contributions. As mentioned earlier,
there are 16 possible NP operators. We present the calculations in
some detail for two of them; the results for the others are included
in tables. We begin with the operator whose coefficient is $f_d^{\sss
RR}$ [Eq.~(\ref{NPoperators})]:
\beq {4 G_{\sss F} \over \sqrt{2}} f_d^{\sss RR} \, {\bar b}
\gamma_{\sss R} s \, {\bar d} \gamma_{\sss R} d ~. \eeq
Because this is a scalar/pseudoscalar operator, within factorization
it does not contribute to $B^+ \to \rho^0 K^{*+}$. However, it can
affect $B^+ \to \rho^+ K^{*0}$. To see this, we perform a Fierz
transformation of this operator (both fermions and colors):
\beq -{4 \over N_c} {G_{\sss F} \over \sqrt{2}} f_d^{\sss RR} \,
\left[
  \frac12 \, {\bar b} \gamma_{\sss R} d \, {\bar d} \gamma_{\sss R} s
  + \frac18 \, {\bar b} \sigma^{\mu\nu} \gamma_{\sss R} d \, {\bar d}
  \sigma_{\mu\nu} \gamma_{\sss R} s \right] ~.
\label{tensor} \eeq
It is the second term which is important (in contrast to $B\to\pi
K$), as it contributes to $B^+ \to \rho^+ K^{*0}$.

Within factorization, the contribution to $B^+ \to \rho^+ K^{*0}$
is given by
\beq -{1 \over 2 N_c} {G_{\sss F} \over \sqrt{2}} f_d^{\sss RR} \,
\bra{K^{*0}} {\bar d} \sigma_{\mu\nu} \gamma_{\sss R} s \ket{0}
\bra{\rho^+} {\bar b} \sigma^{\mu\nu} \gamma_{\sss R} d \ket{B^+}
~. \eeq
Using the matrix elements given in the Appendix, this gives
\bea & & Z_d^{\sss RR} \left\{ 2 T_2
\left( 1-
 { m_{\sss \rho}^2 \over m_{\sss B}^2 } \right)
 \left( \epsilon^*_\rho \cdot
\epsilon^*_{\sss K^*} \right) - {4 \over m_{\sss B}^2} \left( T_2
+ T_3 { m_{\sss K^*}^2 \over m_{\sss B}^2 } \right) \left(
\epsilon^*_\rho \cdot p_{\sss K^*} \right) \left( \epsilon^*_{\sss
K^*} \cdot p_\rho \right) \right. \nn\\
& & \hskip2truecm \left. -~{4 i\over m_{\sss B}^2} T_1
\epsilon^{\mu\nu\alpha\beta} p^\rho_\mu p^{\sss K^*}_\nu
\epsilon^{* \rho}_\alpha \epsilon^{* {\sss K^*}}_\beta \right\} ~,
\eea
where the $T_i$ are form factors and
\beq Z_d^{\sss RR} \equiv {1 \over 4 N_c} {G_{\sss F} \over
\sqrt{2}} f_d^{\sss RR} g_{\sss T}^{\sss K^*} m_{\sss B}^2 ~.
\label{Zdef} 
\eeq

We again use the form factor relations \cite{formfactors}
\bea
 T_{1}(q^2) & \approx & \zeta_\perp ~, \nonumber\\
 T_{2}(q^2) & \approx & \zeta_\perp
\left ( 1-{ q^2 \over {m_{\sss B}^2-m_{\sss V}^2}} \right) ~, \nonumber\\
 T_{3}(q^2) & \approx &
\zeta_\perp - 
{2 m_{\sss V} \over m_{\sss B}} \zeta_\| ~.
\eea
Comparing the above expression for the NP amplitude with the formula
in Eqs.~(\ref{BVVamp}), we see that the NP operator whose coefficient
is $f_d^{\sss RR}$ predicts
\beq 
A_0 = - 2 \zeta_\| {m_{\sss K^*} \over m_{\sss B}} Z_d^{\sss RR} ~,~~
A_\| = 2 \sqrt{2} \zeta_\perp Z_d^{\sss RR} ~,~~ 
A_\perp = 2 \sqrt{2} \zeta_\perp
Z_d^{\sss RR} ~. 
\eeq 
(We note that $A_0$ above is subleading in $1/m_{\sss B}$ and so we
have used the general expressions in Eq.~\ref{Aidefs} instead of
Eq.~\ref{smallmass} to calculate the longitudinal polarization
amplitude.) We therefore see that this operator contributes
significantly to transverse polarization states of $\rho^- K^{*0}$.
The longitudinal polarization is suppressed by $O(m_{\sss V}/m_{\sss
B})$ as expected.

We can now calculate the ratio of transverse and longitudinal
polarizations, including the SM contribution [Eq.~(\ref{SMcont})].
Assuming $g_{K^*} \approx g_{K^*}^{\sss T}$ and taking the value of
the NP coefficient from $B\to\pi K$ [Eq.~(\ref{coeffvalues})], we have
with $T= \perp, \|$
\beq
{ f_{\sss T} \over f_{\sss L}} = 2{ \left|f_d^{\sss RR} / (2 N_c)
\right|^2 \over \left|X \right|^2} = \cases{ 0.22, & $m_d = 4$ MeV,
\cr 0.86, & $m_d = 8$ MeV.  \cr}
\label{ToverLNP}
\eeq
We therefore see that this NP operator can generate a large transverse
polarization in $B^+ \to \rho^+ K^{*0}$.

Note that we also predict for this NP operator (as well as the
operator associated with $f^{\sss LL}$)
\bea
{ f_{\sss \perp} \over f_{\sss \|}} \approx 1 + O(m_{\sss V}/m_{\sss B})\
\label{scalartrans}
\eea
which is the same as the SM prediction.

The second NP operator for which we explicitly present calculations is
the one whose coefficient is $g_u^{\sss LR}$
[Eq.~(\ref{NPoperators})]:
\beq {4 G_{\sss F} \over \sqrt{2}} g_u^{\sss LR} \, {\bar b}
\gamma^\mu \gamma_{\sss L} s \, {\bar u} \gamma_\mu \gamma_{\sss
R} u ~. \eeq
This operator contributes directly to $B^+ \to \rho^0 K^{*+}$. Its
Fierz transformation has the form $(S-P)\times (S+P)$ and, being a
scalar/pseudoscalar operator, does not contribute to $B ^-\to \rho^-
K^{*0}$ within factorization. In this case, the situation is much like
the SM, and using the matrix elements found in the Appendix, the
amplitude corresponding to this operator for $B^+ \to \rho^0 K^{*+}$
is dominantly longitudinal, with
\beq A_0 \simeq {1\over\sqrt{2}} X_u^{\sss LR} \zeta_\| ~~,~~~~
X_u^{\sss LR} \equiv {G_{\sss F} \over \sqrt{2}} g_u^{\sss LR}
g_\rho m_{\sss B}^2 ~. \label{Xdef} \eeq

The contributions of all 16 new-physics operators to the $B\to\rho
K^*$ polarization states are shown in Tables \ref{rho0decays} and
\ref{rho+decays}. Here we present only the dominant contributions to
$f_{\sss L}$ and $f_{\sss T}$; terms of $O(m_{\sss V}/m_{\sss B})$ are
subdominant and contribute to $f_{\sss L,T}$ only at the $O(m_{\sss
V}^2/m_{\sss B}^2)\sim 5$\% level. Of all the operators, there are
only two which reproduce the data of Table \ref{BrhoKstardata}, i.e.\
they contribute significantly to the transverse polarization of $B^+
\to \rho^+ K^{*0}$ while leaving $B^+ \to \rho^0 K^{*+}$ essentially
longitudinal. They have the coefficients $f_d^{\sss RR}$ and
$f_d^{\sss LL}$. These are the only two NP operators which
successfully explain both the $B\to\pi K$ and $B^+\to\rho K^*$ data.

\begin{table}
\caption{Contributions to the polarization states of $B^+ \to \rho^0
K^{*+}$ from the various NP operators. Operators which are not shown
do not contribute. The various $Z$'s and $X$'s are defined analogously
to Eqs.~(\ref{Zdef}) and (\ref{Xdef}). We take $\zeta_\perp \approx
\zeta_\|$.}
\begin{center}
\begin{tabular}{cccc}
\hline \hline
 & $A_0$ & $A_\|$ & $A_\perp$ \\
\hline
$f_u^{\sss RR}$ & $O(m_{\sss V}/m_{\sss B})$ & $2 \zeta_{\perp\rho}   Z_u^{\sss
RR}$ & $2 \zeta_{\perp\rho}   Z_u^{\sss RR}$ \\
$f_u^{\sss LL}$ & $O(m_{\sss V}/m_{\sss B})$ & $-2 \zeta_{\perp\rho}  
Z_u^{\sss LL}$ & $2 \zeta_{\perp\rho}   Z_u^{\sss LL}$ \\
$f_u^{\sss RL}$ & $ -\sqrt{2} \zeta_{\| \rho} (g_{\sss K^*} / g_{\sss
K^*}^{\sss T} ) Z_u^{\sss RL}$ & $O(m_{\sss V}/m_{\sss B})$ &
$O(m_{\sss V}/m_{\sss B})$ \\
$f_u^{\sss LR}$ & $ \sqrt{2} \zeta_{\| \rho} (g_{\sss K^*} g_{\sss
K^*}^{\sss T} ) Z_u^{\sss LR}$ & $O(m_{\sss V}/m_{\sss B})$ &
$O(m_{\sss V}/m_{\sss B})$ \\
$g_u^{\sss RR}$ & $ -{1\over\sqrt{2}} (\zeta_{\| \sss K^*} + (g_{\sss
  K^*} / g_{\sss \rho})\zeta_{\| \sss \rho} / N_c) X_u^{\sss RR} $ &
  $O(m_{\sss V}/m_{\sss B})$ & $O(m_{\sss V}/m_{\sss B})$ \\
$g_u^{\sss LL}$ & $ {1\over\sqrt{2}} (\zeta_{\| \sss K^*} + (g_{\sss
  K^*} / g_{\sss \rho} )\zeta_{\| \sss \rho} / N_c) X_u^{\sss LL} $ &
  $O(m_{\sss V}/m_{\sss B})$ & $O(m_{\sss V}/m_{\sss B})$ \\
$g_d^{\sss RR}$ & ${1\over\sqrt{2}} \zeta_{\| \sss K^*} X_d^{\sss RR}
$ & $O(m_{\sss V}/m_{\sss B})$ & $O(m_{\sss V}/m_{\sss B})$ \\
$g_d^{\sss LL}$ & $-{1\over\sqrt{2}} \zeta_{\| \sss K^*} X_d^{\sss LL}
$ & $O(m_{\sss V}/m_{\sss B})$ & $O(m_{\sss V}/m_{\sss B})$ \\
$g_u^{\sss RL}$ & $-{1\over\sqrt{2}} \zeta_{\| \sss K^*} X_u^{\sss RL}
$ & $O(m_{\sss V}/m_{\sss B})$ & $O(m_{\sss V}/m_{\sss B})$ \\
$g_u^{\sss LR}$ & ${1\over\sqrt{2}} \zeta_{\| \sss K^*} X_u^{\sss LR}
$ & $O(m_{\sss V}/m_{\sss B})$ & $O(m_{\sss V}/m_{\sss B})$ \\
$g_d^{\sss RL}$ & ${1\over\sqrt{2}} \zeta_{\| \sss K^*} X_d^{\sss RL}
$ & $O(m_{\sss V}/m_{\sss B})$ & $O(m_{\sss V}/m_{\sss B})$ \\
$g_d^{\sss LR}$ & $-{1\over\sqrt{2}} \zeta_{\| \sss K^*} X_d^{\sss LR}
$ & $O(m_{\sss V}/m_{\sss B})$ & $O(m_{\sss V}/m_{\sss B})$ \\ \hline
\end{tabular}
\end{center}
\label{rho0decays}
\end{table}

\begin{table}
\caption{Contributions to the polarization states of $B^+ \to \rho^+
K^{*0}$ from the various NP operators. Operators which are not shown
do not contribute. The various $Z$'s and $X$'s are defined analogously
to Eqs.~(\ref{Zdef}) and (\ref{Xdef}). We take $\zeta_\perp \approx
\zeta_\|$.}
\begin{center}
\begin{tabular}{cccc}
\hline \hline
Operator & $A_0$ & $A_\|$ & $A_\perp$ \\
\hline 
$f_d^{\sss RR}$ & $O(m_{\sss V}/m_{\sss B})$ & $2 \sqrt{2}  \zeta_{\perp\rho}  
Z_d^{\sss RR}$ & $2 \sqrt{2} \zeta_{\perp\rho}   Z_d^{\sss RR}$ \\
$f_d^{\sss LL}$ & $O(m_{\sss V}/m_{\sss B})$ & $-2 \sqrt{2} \zeta_{\perp\rho}  
Z_d^{\sss LL}$ & $2 \sqrt{2} \zeta_{\perp\rho}   Z_d^{\sss LL}$ \\
$f_d^{\sss RL}$ & $-2 \zeta_{\| \rho} (g_{\sss K^*} / g_{\sss
K^*}^{\sss T} ) Z_d^{\sss RL}$ & $O(m_{\sss V}/m_{\sss B})$ &
$O(m_{\sss V}/m_{\sss B})$ \\
$f_d^{\sss LR}$ & $ 2 \zeta_{\| \rho} (g_{\sss K^*} / g_{\sss
K^*}^{\sss T} ) Z_d^{\sss LR}$ & $O(m_{\sss V}/m_{\sss B})$ &
$O(m_{\sss V}/m_{\sss B})$ \\
$g_d^{\sss RR}$ & ${1\over N_c} \zeta_{\| \rho} (g_{\sss K^*} /
g_{\rho}) X_d^{\sss RR} $ & $O(m_{\sss V}/m_{\sss B})$ & $O(m_{\sss
V}/m_{\sss B})$ \\
$g_d^{\sss LL}$ & $-{1\over N_c} \zeta_{\| \rho} (g_{\sss K^*} /
 g_{\rho}) X_d^{\sss LL} $ & $O(m_{\sss V}/m_{\sss B})$ & $O(m_{\sss
 V}/m_{\sss B})$ \\
\hline
\end{tabular}
\end{center}
\label{rho+decays}
\end{table}

This explanation of the $B \to \rho K^*$ data can be tested. In the
SM, there is essentially only one dynamical decay amplitude.  Because
of this, one expects the CP-violating triple-product correlation (TP)
in these decays to be very small \cite{BVVTP}.  However, this can
change with the addition of a second NP amplitude. A nonzero value of
the $f_d^{\sss RR}$ or $f_d^{\sss LL}$ amplitude will lead to a
nonzero TP. Furthermore, one expects such a TP only in $B^+ \to \rho^+
K^{*0}$; the TP should remain tiny in $B^+ \to \rho^0 K^{*+}$.

We can estimate the expected size of the TP in $B^+ \to \rho^+
K^{*0}$. In Ref.~\cite{BVVTP} the following measures of the
triple-product correlations were defined:
\beq
A_T^{(1)} \equiv \frac{{\rm Im}(A_\perp
A_0^*)}{A_0^2+A_\|^2+A_\perp^2} ~~,~~~~
A_T^{(2)} \equiv \frac{{\rm Im}(A_\perp
A_\|^*)}{A_0^2+A_\|^2+A_\perp^2} ~.
\label{TPmeasure}
\eeq
The corresponding quantities for the charge-conjugate process,
$\bar{A}_T^{(1)}$ and $\bar{A}_T^{(2)}$, are defined similarly. The
comparison of the TP asymmetries in a decay and in the corresponding
CP-conjugate process will give a measure of the true T-odd,
CP-violating asymmetry for that decay. The TP is therefore due to the
interference between the $A_\perp$ and $A_0$ or $A_\|$ amplitudes, and
requires that the two interfering amplitude have different weak
phases. Recall that it was found in Ref.~\cite{BKpidecays} that, to
explain the $B \to \pi K$ data, a NP weak phase $\phi_{\sss NP} \sim
100^\circ$ was needed.

Now, at leading order, the SM yields only $A_0$; large transverse
amplitudes can arise only if NP is included. However, the only way to
obtain a nonzero $A_T^{(2)}$ is through SM--NP interference. We
observe from Table~\ref{rho+decays} that the NP operators associated
with the coefficients $f_d^{\sss RR}$ and $f_d^{\sss LL}$ yield large
values for $A_\perp$ or $A_\|$. On the other hand, the transverse SM
amplitudes are all $O(1/m_{\sss B})$. Thus, the SM--NP interference
gives an $A_T^{(2)}$ of $O(1/m_{\sss B})$. Note that a measurement of
the sign of $A_T^{(2)}$, if possible, can be used to distinguish
between the two NP operators.

In contrast, the TP asymmetry $A_T^{(1)}$ can be sizeable. It can
arise due to the interference of the $A_0$ SM amplitude and the
$A_{\perp}$ NP amplitude. As above, this latter amplitude can be big
for those NP operators whose coefficients are $f_d^{\sss RR}$ or
$f_d^{\sss LL}$. For these operators, we can estimate the maximum
magnitude of $A_T^{(1)}$. We first take the strong-phase difference
between $A_0$ and $A_{\perp}$ to be zero (or $\pi $). In this case,
$A_T^{(1)}$ is by itself a measure of T-odd CP violation and we can
write
\bea
|A_T^{(1)}| \le \frac{ \sqrt{f_{\perp} / f_{\sss L}}}
{1+ {2f_{\perp} / f_{\sss L}}} \, \sin \phi_{\sss NP} ~.
\eea 
Using $ \phi_{\sss NP} \sim 100^\circ$ and Eq.~\ref{ToverLNP} we find
$|A_T^{(1)}| \le$ 32-34\% for $m_d=$ 4-8 MeV.  Hence we see that a
sizeable TP is possible in the decay $B^+ \to \rho^0 K^{*+}$.
 
\section{$B\to\phi K^*$}
\label{btophik}

As noted earlier, a sizeable value of $f_{\sss T}/f_{\sss L}$ is
observed in $B\to\phi K^*$, contrary to expectations. There are
different SM explanations, but they all predict either that (i) the
transverse polarization fractions are large in both $B^+ \to \rho^+
K^{*0}$ and $B^+ \to \rho^0 K^{*+}$, with the $f_{\sss T}$'s
respecting Eq.~(\ref{fTreln}), or (ii) $f_{\sss T}$ is small in both
$B\to\rho K^*$ decays. If either of these is not seen, new physics is
needed.

There are already several non-SM explanations of the $\phi K^*$ data
\cite{phiKstarNP,DasYang}, but one can now ask the question: can one
explain the $\pi K$, $\rho K^*$ and $\phi K^*$ observations
simultaneously?  The answer is yes. One can reproduce the $\phi K^*$
data with the addition of NP operators of the form ${\bar b}
\gamma_{\sss R} s \, {\bar s} \gamma_{\sss R} s$ or ${\bar b}
\gamma_{\sss L} s \, {\bar s} \gamma_{\sss L} s$ \cite{DasYang}.
Above, we have shown that NP operators such as ${\bar b} \gamma_{\sss
R} s \, {\bar d} \gamma_{\sss R} d$ or ${\bar b} \gamma_{\sss L} s \,
{\bar d} \gamma_{\sss L} d$ can account for the observations in the
$\pi K$ and $\rho K^*$ systems. Thus, if the NP obeys an approximate
U-spin symmetry, which relates $d$- and $s$-quarks, one can
simultaneously explain the $\pi K$, $\rho K^*$ and $\phi K^*$
observations. (A model which does this will be described in
Ref. \cite{Datta}.)

\section{Conclusions}
\label{conclusions} 

At present, there are several discrepancies with the predictions of
the standard model (SM), in $B\to\phi K$, $B\to \phi K^*$ and $B\to
\pi K$ decays. We must stress that these discrepancies are (almost)
all in the 1--2$\sigma$ range and as such are not yet statistically
significant. That is, the existence of physics beyond the SM is not
certain. However, if these hints are taken together, the statistical
significance increases. Furthermore, they are intriguing since they
all point to new physics (NP) in $\btos$ transitions. For these
reasons, it is worthwhile considering the effects of NP on various
$B$ decays.

One hint of NP occurs in the decays $B\to\phi K^*$. The SM naively
predicts that the transverse polarization fraction of the final-state
particles, $f_{\sss T}$, should be much smaller [$O(m_{\sss
V}^2/m_{\sss B}^2)$] than that of the longitudinal polarization,
$f_{\sss L}$. However, it is observed that $f_{\sss T} \simeq f_{\sss
L}$.  There are several SM explanations, all of which go beyond the
naive expectations. However, all make predictions for the polarization
in $B\to\rho K^*$ decays. The key point is that there are two such
decays, $B^+ \to \rho^+ K^{*0}$ and $B^+ \to \rho^0 K^{*+}$ (and
similarly for neutral $B$ decays). By measuring the polarizations in
{\it both} decays, one can test the SM explanations of the $B\to\phi
K^*$ measurements.

In one scenario \cite{colangelo,Kagan}, it is predicted that $f_{\sss
T}$ should be large in both $B\to\rho K^*$ decays. We have shown that
the values of $f_{\sss T}$ in both decays should obey
Eq.~(\ref{fTreln}).  If this relation is not respected, then this
scenario is ruled out, yielding a clear signal of new physics. Using
present $B\to\rho K^*$ data, the central values violate this
relation. However, the errors are still extremely large, so that no
firm conclusions can be drawn. This emphasizes the importance of more
precise measurements of these decays.

In the second scenario \cite{phiKstarSMHou}, the transverse
polarizations in both $B\to\rho K^*$ decays are predicted to be small,
i.e.\ $f_{\sss L}$ is close to 1.  However, in $B^+ \to \rho^+ K^{*0}$
decays, it is found that $f_{\sss L}^+ = 0.66 \pm 0.07$
(Table~\ref{BrhoKstardata}), ruling out this scenario at the
3.5$\sigma$ level.

The discrepancy in $B\to \pi K$ decays can be explained by the
addition of new-physics operators of the form $\btos q{\bar q}$
($q=u,d$) \cite{BKpidecays1, BKpidecays}. There are 16 such operators,
all of which will contribute to $B\to\rho K^*$ decays. Assuming that
NP is present, we have calculated the effect on the polarization
states of $B\to\rho K^*$ of each of these operators (Tables
\ref{rho0decays} and \ref{rho+decays}). Of these, there are only two
which reproduce the data of Table \ref{BrhoKstardata}, i.e.\ they
contribute significantly to the transverse polarization of $B^+ \to
\rho^+ K^{*0}$ while leaving $B^+ \to \rho^0 K^{*+}$ essentially
longitudinal. They are $f_d^{\sss RR} \, {\bar b} \gamma_{\sss R} s \,
{\bar d} \gamma_{\sss R} d$ and $f_d^{\sss LL} \, {\bar b}
\gamma_{\sss L} s \, {\bar d} \gamma_{\sss L} d$. If the $B\to \pi K$
measurements turn out to show statistically-significant evidence of
new physics, and if the $B\to\rho K^*$ data remain as in Table
\ref{BrhoKstardata}, these are the only two NP operators which can
explain both sets of observations.

Finally, it is natural to assume that the same type of new physics
which accounts for the $B\to \pi K$ and $B\to\rho K^*$ measurements
also affects $B\to\phi K^*$ decays and can explain the observed value
of $f_{\sss T}/f_{\sss L}$. This is possible if the NP obeys an
approximate U-spin symmetry. In this case, there are also NP operators
of the form ${\bar b} \gamma_{\sss R} s \, {\bar s} \gamma_{\sss R} s$
or ${\bar b} \gamma_{\sss L} s \, {\bar s} \gamma_{\sss L} s$, which
can reproduce the $\phi K^*$ data \cite{DasYang}. This type of NP can
therefore simultaneously account for the $\pi K$, $\rho K^*$ and $\phi
K^*$ data.

Note that it is quite possible that, with more data, the experimental
measurements will change, leading to a different pattern of
new-physics signals. In this case, the conclusions presented in this
paper will have to be modified. However, we must stress that this type
of analysis will ultimately be necessary. Rather than look for NP
solutions to each individual discrepancy with the SM, it will be far
more compelling to search for a single solution to all NP signals.
Thus, an analysis of the type presented in this paper will have to be
carried out.

\bigskip
\noindent {\bf Acknowledgements}:
This work is financially supported by NSERC of Canada.

\newpage

\section*{Appendix}

The matrix elements used in the paper: we have
\bea \bra{\pi^0} {\bar q} (1 \pm \gamma_5) q \ket{0} & \!\! = \!\!
& \pm i \, {f_\pi
\over \sqrt{2}} \, {m_\pi^2 \over 2m_q} ~, \nn\\
\bra{\pi^0} {\bar q} \gamma^\mu (1 \pm \gamma_5) q \ket{0} & \!\!
= \!\! & \mp i
\, f_{\sss K} p_\pi^\mu ~, \nn\\
\bra{K^+} {\bar u} \gamma^\mu (1 - \gamma_5) s \ket{0} & \!\! =
\!\! & i f_{\sss
K} p_{\sss K}^\mu ~, \nn\\
\bra{K^+} {\bar b} (1 \pm \gamma_5) s \ket{B^+} & \!\! = \!\! &
{m_{\sss B}^2 -
m_{\sss K}^2 \over m_s - m_b} F_0^{\sss K} ~, \nn\\
\bra{K^+} {\bar b} \gamma_\mu (1 \pm \gamma_5) s \ket{B^+} & \!\!
= \!\! & \left[ (p_{\sss B} + p_{\sss K})_\mu - {m_{\sss B}^2 -
m_{\sss K}^2 \over
q^2} \, q_\mu \right] F_1^{\sss K} \nn\\
& & \hskip0.75truecm +~{m_{\sss B}^2 - m_{\sss K}^2 \over q^2} \,
q_\mu F_0^{\sss K} ~, ~~~~ q_\mu \equiv \left( p_{\sss B} -
p_{\sss K}
\right)_\mu ~,\nn\\
\bra{\pi^0} {\bar b} \gamma_\mu (1 - \gamma_5) u \ket{B^+} & \!\!
= \!\! & \left[ (p_{\sss B} + p_\pi)_\mu - {m_{\sss B}^2 - m_\pi^2
\over q^2} \, q_\mu \right] F_1^\pi \nn\\
& & \hskip0.75truecm +~{m_{\sss B}^2 - m_\pi^2 \over q^2} \, q_\mu
F_0^\pi
~, ~~~~ q_\mu \equiv \left( p_{\sss B} - p_\pi \right)_\mu ~, \nn\\
\bra{K^*} {\bar q} \gamma^\mu s \ket{0} & \!\! = \!\! & g_{\sss
K^*} m_{\sss
K^*} \epsilon^{*\mu}_{\sss K^*} ~, \nn\\
\bra{\rho} {\bar b} \gamma^\mu (1 \pm \gamma_5) q \ket{B}
\epsilon^{* \sss K^*}_\mu & \!\! = \!\! & {2 i \over m_{\sss B} +
m_\rho } V^\rho \epsilon^{\mu\nu\alpha\beta} p^\rho_\mu p^{\sss
K^*}_\nu \epsilon^{* \rho}_\alpha \epsilon^{* \sss K^*}_\beta \pm
\left( m_{\sss B} + m_\rho \right) A_1^\rho \epsilon^{* \rho}
\cdot \epsilon^{* \sss K^*}
\nn\\
& & \hskip0.75truecm \mp~ A_2^\rho {2 \over m_{\sss B} + m_\rho }
\left( p^\rho \cdot \epsilon^{* \sss K^*} \right) \left( p^{\sss
K^*}
\cdot \epsilon^{* \rho} \right) ~, \nn\\
\bra{K^*} {\bar q} \sigma^{\mu\nu} s \ket{0} & \!\! = \!\! & - i
\, g_{\sss T}^{\sss K^*} \left( \epsilon^{*\mu}_{\sss K^*} p_{\sss
K^*}^\nu -
\epsilon^{*\nu}_{\sss K^*} p_{\sss K^*}^\mu \right) ~, \nn\\
\bra{\rho} {\bar b} \sigma^{\mu\nu} q \ket{B} p^{\sss K^*}_\nu &
\!\! = \!\! & - 2 T_1 \epsilon^{\mu\nu\alpha\beta} p^{\sss
K^*}_\nu p^\rho_\alpha
\epsilon^{* \rho}_\beta ~, \nn\\
\bra{\rho} {\bar b} \sigma^{\mu\nu} \gamma_5 q \ket{B} p^{\sss
K^*}_\nu & \!\! = \!\! & - i T_2 \left[ \left( m_{\sss B}^2 -
m_\rho^2 \right) \epsilon^{* \mu}_\rho - \left( \epsilon^*_\rho
\cdot p_{\sss K^*}
\right) \left( p_{\sss B}^\mu + p_\rho^\mu \right) \right] \nn\\
& & \hskip0.75truecm -~i T_3 \left( \epsilon^*_\rho \cdot p_{\sss
K^*} \right) \left[ p_{\sss K^*}^\nu - { m_{\sss K^*}^2 \over
m_{\sss B}^2
- m_\rho^2} \left( p_{\sss B}^\mu + p_\rho^\mu \right) \right] ~, \nn\\
 \bra{\rho^0} {\bar u} \gamma^\mu u \ket{0} & \!\! = \!\! &
{1\over\sqrt{2}} g_{\sss \rho} m_{\sss \rho} \epsilon^{*\mu}_{\sss
\rho} ~, \nn\\
\bra{K^*} {\bar b} \gamma^\mu (1 \pm \gamma_5) s \ket{B}
\epsilon^{* \rho}_\mu & \!\! = \!\! & {2 i \over m_{\sss B} +
m_{\sss K^*} } V^{\sss K^*} \epsilon^{\mu\nu\alpha\beta}
p^\rho_\mu p^{\sss K^*}_\nu \epsilon^{* \rho}_\alpha \epsilon^{*
\sss K^*}_\beta \pm \left( m_{\sss B} + m_{\sss K^*} \right)
A^{\sss K^*}_1 \epsilon^{* \rho} \cdot
\epsilon^{* \sss K^*} \nn\\
& & \hskip0.75truecm \mp~ A^{\sss K^*}_2 {2 \over m_{\sss B} +
m_{\sss K^*} } \left( p^\rho \cdot \epsilon^{* \sss K^*} \right)
\left( p^{\sss K^*} \cdot \epsilon^{* \rho} \right) ~. \eea



\begin{thebibliography}{99}

\bibitem{sin2beta} B.~Aubert {\it et al.}  [BABAR Collaboration],
Phys.\ Rev.\ Lett.\ {\bf 94} (2005) 161803; K.~Abe {\it et al.}
[BELLE Collaboration], Phys.\ Rev.\ D {\bf 71} (2005) 072003.

\bibitem{sin2betapeng} K.~F.~Chen [Belle Collaboration],
[hep-ex/0504023]; B.~Aubert {\it et al.}  [BABAR Collaboration],
[hep-ex/0503011]; B.~Aubert {\it et al.}  [BABAR Collaboration],
[hep-ex/0502019]; B.~Aubert {\it et al.}  [BABAR Collaboration],
[hep-ex/0502017].

\bibitem{BVVTP} For a study of triple products in the SM and with new
physics, see A.~Datta and D.~London, Int.\ J.\ Mod.\ Phys.\ A {\bf
19}, 2505 (2004).

\bibitem{phiKstarTP} B.~Aubert {\it et al.}  [BABAR Collaboration],
Phys.\ Rev.\ Lett.\ {\bf 93}, 231804 (2004); K.~Senyo [Belle
Collaboration], arXiv:hep-ex/0505067.

\bibitem{BKpiexp} CLEO Collaboration, A.~Bornheim {\it et al.}, Phys.\
Rev.\ D {\bf 68}, 052002 (2003); CLEO Collaboration, S.~Chen {\it et
al.}, Phys.\ Rev.\ Lett.\ {\bf 85}, 525 (2000); Belle Collaboration,
Y.~Chao {\it et al.}, Phys.\ Rev.\ D {\bf 69}, 111102 (2004);
hep-ex/0407025, Phys.\ Rev.\ Lett.\ {\bf 93}, 191802 (2004); BELLE
Collaboration, K.~Abe {\it et al.}, hep-ex/0409049; BABAR
Collaboration, B.~Aubert {\it et al.}, Phys.\ Rev.\ Lett.\ {\bf 89},
281802 (2002), hep-ex/0408062, hep-ex/0408080, hep-ex/0408081, Phys.\
Rev.\ Lett.\ {\bf 93}, 131801 (2004).

\bibitem{BKpidecays1} A.~J.~Buras, R.~Fleischer, S.~Recksiegel and
F.~Schwab, Eur.\ Phys.\ J.\ C {\bf 32}, 45 (2003), Phys.\ Rev.\ Lett.\
{\bf 92}, 101804 (2004), Nucl.\ Phys.\ B {\bf 697}, 133 (2004), Acta
Phys.\ Polon.\ B {\bf 36}, 2015 (2005); V.~Barger, C.~W.~Chiang,
P.~Langacker and H.~S.~Lee, Phys.\ Lett.\ B {\bf 598}, 218 (2004);
S.~Mishima and T.~Yoshikawa, Phys.\ Rev.\ D {\bf 70}, 094024 (2004);
Y.~L.~Wu and Y.~F.~Zhou, Phys.\ Rev.\ D {\bf 71}, 021701 (2005);
H.~Y.~Cheng, C.~K.~Chua and A.~Soni, Phys.\ Rev.\ D {\bf 71}, 014030
(2005); Y.~Y.~Charng and H.~n.~Li, Phys.\ Rev.\ D {\bf 71}, 014036
(2005); X.~G.~He and B.~H.~J.~McKellar, arXiv:hep-ph/0410098.

\bibitem{BKpidecays} S.~Baek, P.~Hamel, D.~London, A.~Datta and
D.~A.~Suprun, Phys.\ Rev.\ D {\bf 71}, 057502 (2005).

\bibitem{pdg} Particle Data Group Collaboration, S.~Eidelman {\it et
al.}, Phys.\ Lett.\ B {\bf 592} (2004) 1.

\bibitem{phiKstarexp} B.~Aubert {\it et al.}  [BABAR Collaboration],
Phys.\ Rev.\ Lett.\ {\bf 91}, 171802 (2003); K.~F.~Chen {\it et
al.} [BELLE Collaboration], arXiv:hep-ex/0503013.

\bibitem{phiKstarNP} C.~Dariescu, M.~A.~Dariescu, N.~G.~Deshpande and
D.~K.~Ghosh, Phys.\ Rev.\ D {\bf 69}, 112003 (2004); E.~Alvarez,
L.~N.~Epele, D.~G.~Dumm and A.~Szynkman, arXiv:hep-ph/0410096;
Y.~D.~Yang, R.~M.~Wang and G.~R.~Lu, Phys.\ Rev.\ D {\bf 72}, 015009
(2005); C.~S.~Kim and Y.~D.~Yang, arXiv:hep-ph/0412364; C.~H.~Chen and
C.~Q.~Geng, Phys.\ Rev.\ D {\bf 71}, 115004 (2005).
  
\bibitem{DasYang} P.~K.~Das and K.~C.~Yang, Phys.\ Rev.\ D {\bf 71}, 094002
(2005); 

\bibitem{colangelo} P.~Colangelo, F.~De Fazio and T.~N.~Pham, Phys.\
Lett.\ B {\bf 597}, 291 (2004); H.~Y.~Cheng, C.~K.~Chua and A.~Soni,
Ref.~\cite{BKpidecays1}; M.~Ladisa, V.~Laporta, G.~Nardulli and
P.~Santorelli, Phys.\ Rev.\ D {\bf 70}, 114025 (2004).

\bibitem{Kagan}A.~L.~Kagan, Phys.\ Lett.\ B {\bf 601}, 151 (2004).

\bibitem{phiKstarSMHou} W.~S.~Hou and M.~Nagashima,
arXiv:hep-ph/0408007. 

\bibitem{GHLR} M.~Gronau, O.~F.~Hern\'andez, D.~London and
J.~L.~Rosner, Phys.\ Rev.\ D {\bf 50}, 4529 (1994), Phys.\ Lett.\ B
{\bf 333}, 500 (1994), Phys.\ Rev.\ D {\bf 52}, 6374

\bibitem{EWPs} M.~Neubert and J.~L.~Rosner, Phys.\ Lett.\ B {\bf 441},
403 (1998), Phys.\ Rev.\ Lett.\ {\bf 81}, 5076 (1998); M.~Gronau,
D.~Pirjol and T.~M.~Yan, Phys.\ Rev.\ D {\bf 60}, 034021 (1999)
[Erratum-ibid.\ D {\bf 69}, 119901 (2004)]; M.~Imbeault,
A.~L.~Lemerle, V.~Page and D.~London, Phys.\ Rev.\ Lett.\ {\bf 92},
081801 (2004).

\bibitem{BuraseffH} See, for example, G. Buchalla, A.J. Buras and
  M.E. Lautenbacher, {\it Rev.\ Mod.\ Phys.} {\bf 68}, 1125 (1996).

\bibitem{BrhoKstarBRsfL} Particle Data Group, Ref. \cite{pdg};
B.~Aubert [BABAR Collaboration], arXiv:hep-ex/0408093; J.~Zhang {\it
et al.}  [BELLE Collaboration], arXiv:hep-ex/0505039.

\bibitem{HFAG} The Heavy Flavor Averaging Group,
http://www.slac.stanford.edu/xorg/hfag/

\bibitem{pqcd1}H.~n.~Li and S.~Mishima, Phys.\ Rev.\ D {\bf 71},
054025 (2005).

\bibitem{pqcd2}H.~n.~Li, arXiv:hep-ph/0411305.

\bibitem{DLNP} A.~Datta and D.~London, Phys.\ Lett.\ B {\bf 595}, 453
  (2004).

\bibitem{BNPmethods} A.~Datta, M.~Imbeault, D.~London, V.~Page,
N.~Sinha and R.~Sinha, Phys.\ Rev.\ D {\bf 71}, 096002 (2005).

\bibitem{beneke} M. Beneke, G. Buchalla, M. Neubert and
C.T. Sachrajda, Nucl.\ Phys.\ B {\bf 591}, 313 (2000), Nucl.\ Phys.\ B
{\bf 606}, 245 (2001).

\bibitem{formfactors} J.~Charles, A.~Le Yaouanc, L.~Oliver, O.~Pene
and J.~C.~Raynal, Phys.\ Rev.\ D {\bf 60}, 014001 (1999).  
= HEP-PH 9812358;

\bibitem{Datta} A. Datta, in preparation.

\end{thebibliography}
\end{document}